\documentclass[twocolumn]{aastex631}

\usepackage{soul}
\usepackage{xcolor}
\shorttitle{Different periodic behaviours of magnetic helicity flux in Active Regions}
\shortauthors{So\'os et al.}

\graphicspath{{./}}

\begin{document}

\title{On the differences in the periodic behaviour of magnetic helicity flux\\
in flaring active regions with and without X-class events}

\correspondingauthor{Marianna Kors\'os}
\email{komabi@gmail.com}

\author[0000-0002-3606-161X]{Sz. So\'os}
\affiliation{Department of Astronomy, E\"otv\"os Lor\'and University\\
P\'azm\'any P\'eter s\'et\'any 1/A, H-1112 Budapest, Hungary}
\affiliation{Hungarian Solar Physics Foundation,
Pet\H{o}fi t\'er 3, H-5700 Gyula, Hungary}

\author[0000-0002-0049-4798]{M. B. Kors\'os}
\affiliation{Department of Physics, Aberystwyth University\\
Ceredigion, Cymru SY23 3BZ, UK}
\affiliation{Department of Astronomy, E\"otv\"os Lor\'and University\\
P\'azm\'any P\'eter s\'et\'any 1/A, H-1112 Budapest, Hungary}
\affiliation{Hungarian Solar Physics Foundation,
Pet\H{o}fi t\'er 3, H-5700 Gyula, Hungary}

\author[0000-0002-6547-5838]{H. Morgan}
\affiliation{Department of Physics, Aberystwyth University\\
Ceredigion, Cymru SY23 3BZ, UK}

\author[0000-0003-3439-4127]{R. Erd\'elyi}
\affiliation{Solar Physics \& Space Plasma Research Center (SP2RC), School of Mathematics and Statistics\\
University of Sheffield, Hounsfield Road S3 7RH, UK}
\affiliation{Department of Astronomy, E\"otv\"os Lor\'and University\\
P\'azm\'any P\'eter s\'et\'any 1/A, H-1112 Budapest, Hungary}
\affiliation{Hungarian Solar Physics Foundation,
Pet\H{o}fi t\'er 3, H-5700 Gyula, Hungary}

\begin{abstract}

Observational pre-cursors of large solar flares provide a basis for future operational systems for forecasting. Here, we study the evolution of the normalized emergence (EM), shearing (SH) and total (T) magnetic helicity flux components for 14 flaring with at least one X-class flare) and 14 non-flaring ($<$ M5-class flares) active regions (ARs) using the Spaceweather Helioseismic Magnetic Imager Active Region Patches vector magnetic field data. Each of the selected ARs contain a $\delta$-type spot. The three helicity components of these ARs were analyzed using wavelet analysis. Localised peaks of the wavelet power spectrum (WPS) were identified and statistically investigated. We find that: i) the probability density function of the identified WPS peaks for all the EM, SH and T profiles can be fitted with a set of Gaussian functions centered at distinct periods between $\sim$ 3 to 20 hours. ii) There is a noticeable difference in the distribution of periods found in the EM profiles between the flaring and non-flaring ARs, while no significant difference is found in the SH and T profiles. iii) In flaring ARs, the distributions of the shorter EM/SH/T periods ($<$ 10 hrs) split up into two groups after flares, while the longer periods ($>$ 10 hrs) do not change. iv) When the EM periodicity does not contain harmonics, the ARs do not host a large energetic flare. Finally, v) significant power at long periods ($\sim$ 20 hour) in the T and EM components may serve as pre-cursor for large energetic flares.

\end{abstract}


\section{Introduction} \label{sec:intro}

The interaction of solar activity with Earth's atmosphere occurs through a complex series of events called Space Weather (SW).
The energetic solar flares and Coronal Mass Ejections (CMEs) have dominant roles in SW, because they can cause disruption to human technology, e.g. for the functioning of electric power grids, aviation, radio communication, GPS, and space-based facilities \citep{Eastwood_2017}.
For this reason, it is vital to further develop existing prediction capabilities through the identification of observable
precursors of flares and CMEs \citep[see][]{Barnes_2016, Leka_2019, Kusano2020, patsourakos20, ahmadzadeh21, Georgoulis2021}. Understanding the physical processes of
flare and CME precursors is still a challenging task in solar physics research \citep[][and in their references]{Florios_2018, korsos19}.
The most intense solar eruptions originate from the magnetically most complex, and highly twisted $\delta$-type active regions (ARs) \citep{Georgoulis_2019, Toriumi2019}. We employ this working hypothesis, and focus on the observational property of magnetic helicity flux in $\delta$-type ARs in this work. 

The source of magnetic helicity lies below the photosphere, and can be derived from magnetogram observations of the photosphere. Certain properties of the helicity are thought to be promising parameters to describe the pre-flare states of ARs. This is, partially, because the magnetic helicity flux often has a strong gradient before the flare and CME occur \citep{elsasser1956,Moon_2002a, Moon_2002b, Smyrli2010, Park_2008, Park_2012}. The magnetic helicity carries information about the complexity of the magnetic field topology, and is therefore linked to the free magnetic energy of ARs, and the occurrence of flares. Thus diagnostics related to the magnetic helicity may be valuable for flare prediction \citep{pariat_2017,Thalmann_2019,Korsos_2020}.

Recently, \citet{prior2020} showed that the multi-resolution wavelet decomposition is a useful tool to analyze the magnetic helicity.
Based on their theoretical work, \citet{Korsos_2020} investigated the dynamic evolution of emergence (EM), shearing (SH), and total (T) magnetic
helicity flux terms using a wavelet analysis in the case of three flaring and three non-flaring ARs. They found a relationship between the oscillatory behavior of the three magnetic helicity flux components and the associated flare activities. Their conjecture was that the
three helicity flux components have common period(s) before flare onset. In comparison, the non-flaring ARs did not exhibit such common periodicities.

To further test the conjecture of \citet{Korsos_2020}, this work extends their approach by applying a more extensive set of diagnostics on a larger number of ARs. In this work, we investigate 14 flaring and 14 non-flaring ARs, and apply additional statistical tests, e.g.: the Kolmogorov-Smirnov (KS) test, and the Gaussian Mixture Model (GMM). The selection criteria of the studied 28 ARs are listed in Section~\ref{sec:data}. We describe the method for the derivation of the magnetic helicity components in Section~\ref{sec:method}. Finally, Section~\ref{sec:analysis} presents the analysis and summarises the main findings of our work.

\section{Data} \label{sec:data}

In this study, similar to \citet{Korsos_2020}, active regions that contain X-class flares will be called flaring ARs. While active regions that do not contain
X-class flares are defined as non-flaring ARs. As in \citet{Korsos_2020}, a random sample of 14 flaring and 14 non-flaring active regions are selected based on the following criteria: 

\begin{itemize}
  \item the angular distance of an AR from the central meridian is up to $\pm 60\arcdeg$, to obtain the best possible quality data \citep{Bobra_2014}.
  \item the AR must have a $\delta$-spot configuration.
  \item the flaring ARs must be the location of at least one X-class flare.
  \item the non-flaring ARs should not be the host of flares larger than M5. 
  \item the non-flaring ARs cannot be associated with fast CMEs. Here, we define CMEs with speeds 750\,km~s$^{-1}$ or higher as a fast CME.
\end{itemize}

The selected 28 ARs are listed in Table~\ref{tab:ARs}, along with the largest flare intensity during the time of observation. The non-flaring ARs host flares of intensity classes A, B, and C except AR 11542 which produced one M1.6 flare, and AR 11726 which produced one M1.0 flare. The M-class flare intensities are at the boundary of what is considered to be a truly energetic flare.

\begin{deluxetable}{cccc}
\tablecaption{Summary table of the studied 28 active regions\label{tab:ARs}}
\tablewidth{0pt}
\tablehead{
\twocolhead{Flaring active regions} &
\twocolhead{Non-flaring active regions} \\
\colhead{NOAA} &
\colhead{Largest flare} &
\colhead{NOAA} &
\colhead{Largest flare} \\
\colhead{Number} &
\colhead{intensity} &
\colhead{Number} &
\colhead{intensity}
}
\startdata
AR 11158 & X-class & AR 11271 & C-class \\
AR 11166 & X-class & AR 11281 & C-class \\
AR 11283 & X-class & AR 11363 & C-class \\
AR 11429 & X-class & AR 11465 & C-class \\
AR 11430 & X-class & AR 11542 & M1.6-class \\
AR 11515 & X-class & AR 11678 & C-class \\
AR 11520 & X-class & AR 11726 & M1.0-class \\
AR 11890 & X-class & AR 11785 & C-class \\
AR 11944 & X-class & AR 12104 & C-class \\
AR 12017 & X-class & AR 12108 & C-class \\
AR 12158 & X-class & AR 12175 & C-class \\
AR 12192 & X-class & AR 12280 & C-class \\
AR 12297 & X-class & AR 12645 & C-class \\
AR 12673 & X-class & AR 12740 & C-class \\
\enddata
\tablecomments{The flaring ARs hosted at least one X-class flare, while the non-flaring ARs produced flares
below M2.0-class. All selected ARs contain a complex $\delta$-type spot.}
\end{deluxetable}

\section{Calculation of the magnetic helicity flux} \label{sec:method}

Following \citet{Korsos_2020}, we determine the EM, SH, and T helicity flux components, given by the terms of equation \citep[see,][]{Mitchell_1984}:
\begin{equation}
T \equiv \frac{dH}{dt}\bigg|_{S} = 
2\int_{S}(\vec{A}_{p}\cdot\vec{B}_{h})\vec{v}_{\perp z}dS -
2\int_{S}(\vec{A}_{p}\cdot\vec{v}_{\perp h})\vec{B}_{z}dS,
\label{eq:helflux}
\end{equation}
where $\vec{A}_{p}$ is the vector potential of the potential magnetic field $\vec{B}_{p}$. $\vec{B}_{h}$, $\vec{B}_{z}$, $\vec{v}_{\perp h}$ and
$\vec{v}_{\perp z}$ are the tangential and normal components of the magnetic field and the tangential and normal components of velocity, respectively.
On the right hand side of Eq.~\ref{eq:helflux}, the first term comes from the twisted magnetic flux tubes emerging into the solar atmosphere or also 
submerging into the subsurface layer (emerging component, EM). The second term comes from the shearing and braiding of the field lines, which is caused
by the tangential motions on the solar surface (shearing component, SH).

The helicity is estimated for each AR using the Spaceweather Helioseismic and Magnetic Imager Active Region Patches (SHARPs) vector magnetic field
measurements \citep{Bobra_2014} with a cadence of 12 minutes (\textit{hmi.sharp\_cea\_720s}). The data is rebinned spatially by a factor of four reduction in order to make the code run faster.

Additional constraints are applied to mitigate the artificial 12- and 24-hr periods of the SDO/HMI magnetogram measurements. Based on \cite{Smirnova_2013}, the strength of the magnetic field is capped at a maximum of $\left| \pm 2000\right|$ G. We also impose a minimum threshold of $\left| \pm 200\right|$ G, which is commonly used in the literature \citep{Tziotziou_2015}. 

After the data binning and the applied constraints, the photospheric plasma velocity is estimated by applying the Differential Affine Velocity Estimator for Vector Magnetograms (DAVE4VM) algorithm with a 19-pixel window \citep{Schuck_2008}. The vector potential 
$\vec{A}_{p}$ was derived by the multigrid MUDPACK software \citep{ADAMS_1993}, solving the relevant elliptic partial differential equations.

\section{Analysis} \label{sec:analysis}

The wavelet power spectrum (WPS) is applied to both original and smoothed time series of EM, SH, T. Similar to \citet{Korsos_2020}, the smoothed series are subject to a smoothing window of 24 hrs that was subtracted from the original data in order to further reduce the 12 and 24 hrs SDO artifact \citep{Smirnova_2013}. Following \citet{Korsos_2020}, we employ the WPS algorithm of \citet{Torrence_1998}, using the default Morlet wavelet. From the WPS, global power
spectra (GPS) are calculated through averaging the WPS over time. We identify significance in regions of the WPS based on a $1\sigma$ level, estimated using a white noise model and the standard deviation of the input signal. 

\begin{figure*}
\plotone{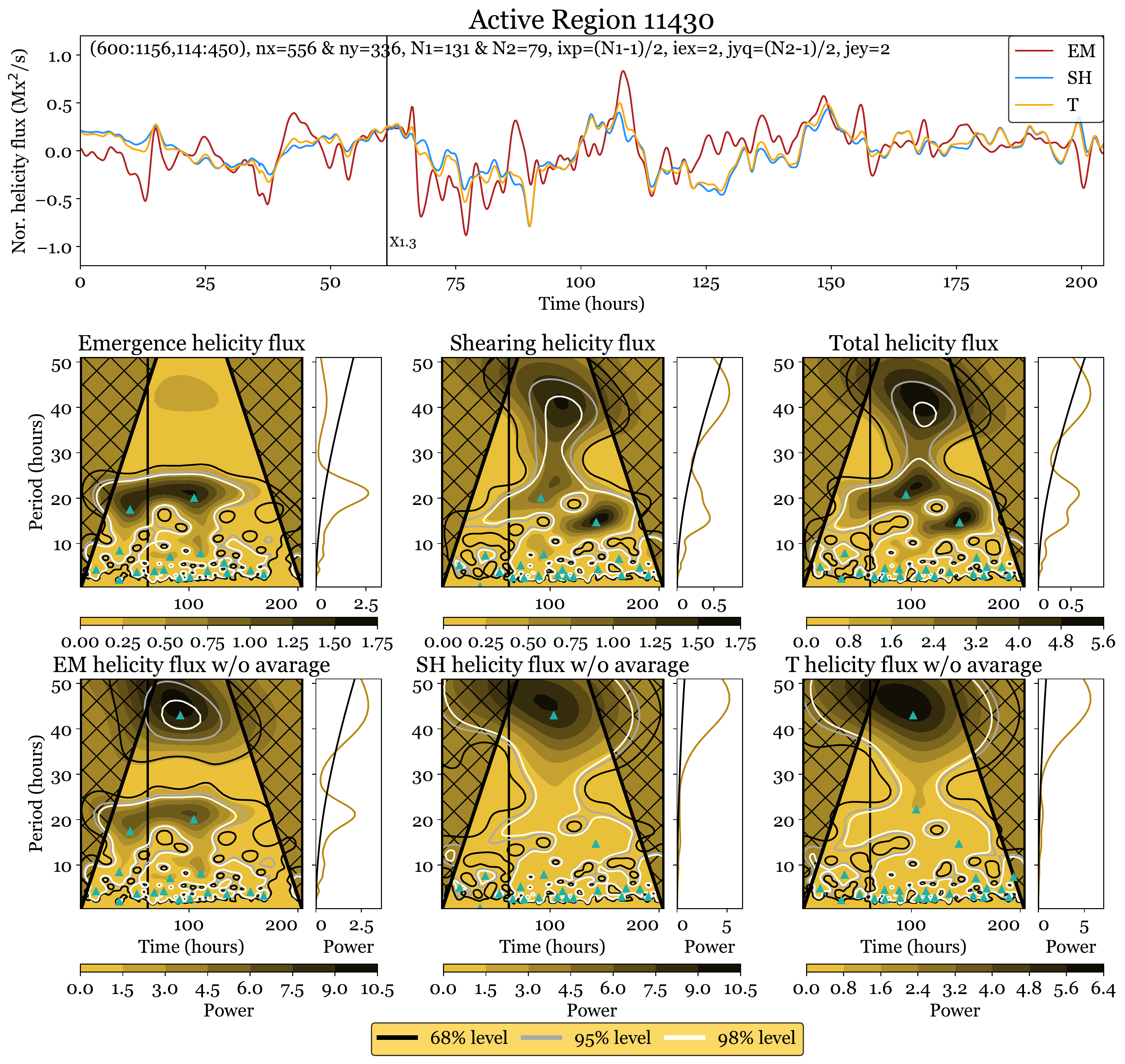}
\caption{Wavelet analysis of the flaring AR 11430. Black vertical lines in all plots mark the on-set time of the X1.3 flare. The top panel shows the unsmoothed EM/SH/T time series in red/blue/orange, respectively. The middle row panels display the WPS of the smoothed EM/SH/T helicity, while the bottom row panels depict the WPS of the unsmoothed time series. For each wavelet, the contour lines mark the significance levels (black at 68\%, gray at 95\%, and white at 98\%), while the shaded color corresponds to the power, as shown in the color bars. In the WPS plots, the thick black lines bounding the gridded regions show the cone-of-influence, i.e., the domain where edge effects become important. Peaks of local maxima within significant regions, identified using a homology method (see text), are shown by the cyan-coloured triangles. The plots to the right of each WPS are the GPS, or the time-averaged WPS. \label{fig:wpspeaks}}
\end{figure*}

 In contrast with \citet{Korsos_2020}, here, we identify local maxima in the WPS by using an implementation of the 0th dimensional persistent homology method \citep{Huber_2021} from a Python package\footnote{\url{http://git.sthu.org/}}. It is important to note that we sought for peaks for significance levels and not for power. These peaks are shown as cyan-coloured triangles for the case of AR 11430 in Figure~\ref{fig:wpspeaks}. Only significant peaks inside the cone-of-influence are considered. These peaks are recorded for all 28 ARs, for both the original and smoothed series of all three helicity flux components. As expected due to a consideration of the wavelet decomposition at different periods, there are larger numbers of peaks at shorter periods (between 1 and 10 hours) whilst only a few peaks are detected at longer periods.

\subsection{Periodicity distribution of significant peaks} \label{sec:peaksditributions}

\begin{figure*}[!th]
\plotone{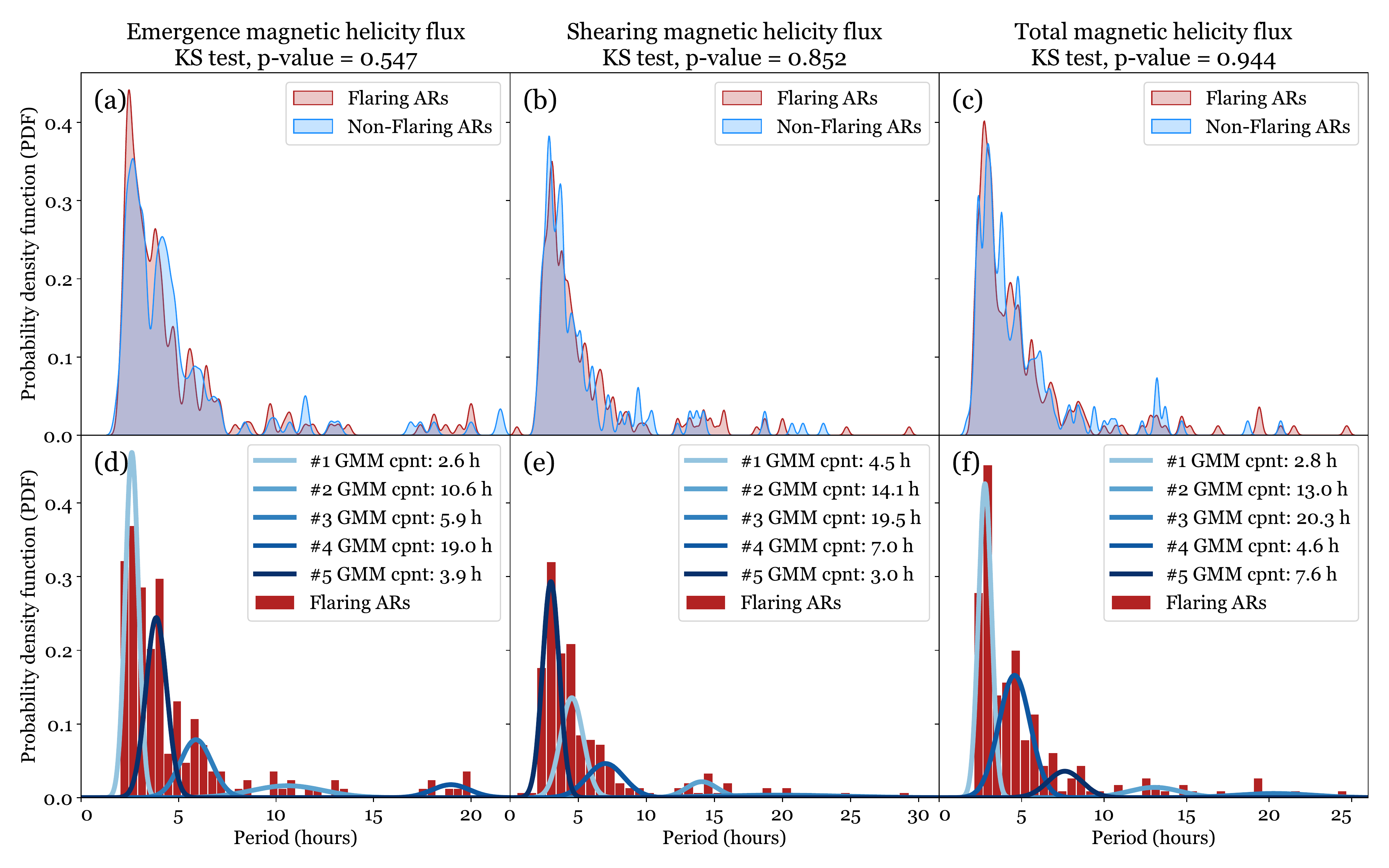}
\caption{Top row: Normalized PDF of the WPS peaks of flaring (red) and non-flaring (blue) ARs for (a) emergence, (b) shearing, and (c) total helicity, using KDE. Note these results are for the smoothed time series. Using GMM, panels d-f show the corresponding distributions fitted to a set of Gaussian functions. The number of Gaussian components are set to 5. The $1/n$-dependence refers to a power law, where $n=1, 2, 3, ...$ is a positive integer, of the EM peaks of Gaussians is clearly visible in panel d. \label{fig:fig2}}
\end{figure*}

The distribution of the identified WPS peaks are analysed with a Kernel Density Estimation (KDE) method. Fig.~\ref{fig:fig2} visualises the KDE analysis of the smoothed EM/SH/T data and Figs.~\ref{fig:fig2}d-f show the GMM results. 

In particular, Figs. \ref{fig:fig2}a-c show the normalized Probability Density Function (PDF) of the identified peaks as a function of periodicity. The PDF of flaring ARs are higher at shorter periods (0-10 hrs) than the non-flaring ones for the EM (Fig.~\ref{fig:fig2}a), and lower for the SH (Fig.~\ref{fig:fig2}b). Based on Fig.~\ref{fig:fig2}a, the magnetic helicity flux EM indicates some level of differences for flaring vs. non-flaring ARs, with \textit{p} = 0.547.
But, we also need to note that {p} = 0.547 is still large and not statistically significant to reject the null hypothesis that the flaring and non-flaring ARs have the same distribution of the WPS peaks. 
However, the distribution of the WPS peaks are the same in the case of SH and T, respectively, with the \textit{p}-values of 0.852 and 0.944. The PDFs of Fig.~\ref{fig:fig2}a-c show bands of preferred periodicities between (i) 2--9 hours,
(ii) 11--14 hours, and (iii) 19--21 hours. We propose that these bands could indicate some global harmonic properties of the EM, SH, and T fluxes. To test this hypothesis, a Kolmogorov-Smirnov (KS) test was performed to compare the peak distributions of flaring and non-flaring ARs, with the results summarized in Table~\ref{tab:allKStest}.

While performing the KS analysis, our null hypothesis was that the similar periods are generated by the same driving mechanism for both flaring and non-flaring ARs. From Table~\ref{tab:allKStest}, we see that all of the \textit{p}-values are below $1\sigma$, which means that the null hypothesis
is true. At this stage, we do not know what exactly is the background driving mechanism for these oscillatory behaviors in the peak distributions of
the three helicity fluxes. Therefore, since the null hypothesis is correct, we can assume that flaring and non-flaring ARs have the same global
harmonic properties.

Let us next focus on the KS test results of the T flux. It is not possible to distinguish between the two cases, namely, flaring vs. non-flaring ARs.
It is worth noting that there is almost no change in the outcome of the KS test results applied to T flux whether using the original or smoothed data.
On the other hand, the results of the KS test of the SH (Fig.~\ref{fig:fig2}b) and the EM (Fig.~\ref{fig:fig2}a) fluxes do reveal different period bands
in the flaring and non-flaring ARs.

\begin{deluxetable}{rcccc}
\tablecaption{KS test results for flaring vs. non-flaring ARs peaks\label{tab:allKStest}}
\tablewidth{0pt}
\tablehead{
\colhead{Helicity flux} &
\twocolhead{with average} &
\twocolhead{w/o average} \\
\colhead{} &
\colhead{Statistic} &
\colhead{\textit{p}-value} &
\colhead{Statistic} &
\colhead{\textit{p}-value}
}
\startdata
Emergence (EM) & 0.086 & 0.547 & 0.048 & 0.991 \\ 
Shearing (SH) & 0.063 & 0.852 & 0.097 & 0.365 \\
Total (T) & 0.054 & 0.944 & 0.054 & 0.942 \\
\enddata
\end{deluxetable}

\subsection{Modelling the PDFs with GMM} \label{sec:pdfsgmm}

Let us now fit the PDFs of the original and smoothed EM/SH/T fluxes, for the flaring and non-flaring cases, with Gaussians in order to reveal any regularity in the peak distributions.
Namely, in Figs. \ref{fig:fig2}d-f, the Gaussian fits are performed by employing the GMM. At first, the number of fitted Gaussian distributions were determined by the best Akaike Information Criterion (AIC) and Bayesian Information Criterion (BIC) values, which supported our earlier findings that the periods are aggregated in bands. The AIC/BIC analysis identified 5 components for the EM of flaring ARs, and only 2 or 3 components for the SH and T. In order to make a consistent comparison between all time series, and from visual inspection of Figs.~\ref{fig:fig2}d-f, we impose a set number of five GMM components for our analysis. 

The central periodicities of the fitted GMM are summarized in Table~\ref{tab:5gmm}. In each case, the fundamental frequency belongs to the largest period. Based on this fundamental frequency, we also calculated the expected higher harmonics, which are shown in square brackets in Table~\ref{tab:5gmm}. In Fig. \ref{fig:fig2}d the
$1/n$-dependence (n=1,2,3,...) of the GMM-fitted Gaussian peaks are, again, clearly visible, which may serve to be evidence for the presence of global eigenmodes. 

\begin{figure}[!ht]
\plotone{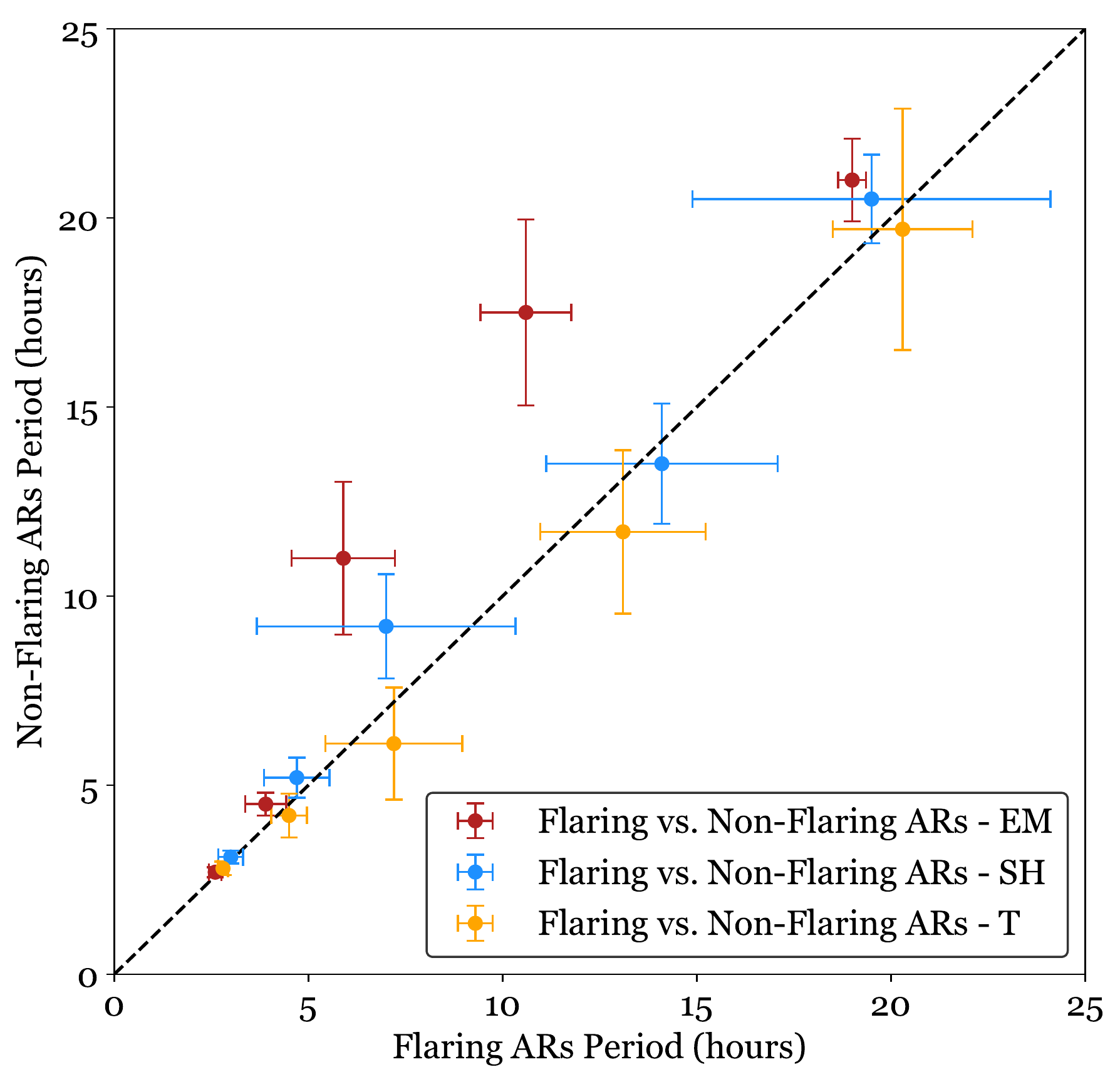}
\caption{Correlation between the Gaussian-central periods of flaring and non-flaring ARs. The error bars are estimated with a bootstrapping method. The dashed black line shows the one-to-one correspondence between flaring and non-flaring periodicities. Only the EM periodicities are significantly above this line. \label{fig:peakscorr}}
\end{figure}

Next, we investigate the relationship between the Gaussian central periods obtained by GMM of the flaring and non-flaring ARs. This relationship is visualised in Fig.~\ref{fig:peakscorr}, where the \textit{x}-axis shows the periods of flaring ARs while the \textit{y}-axis is the periods of non-flaring ARs.
The errors of the mean periods are estimated for the obtained periods using a bootstrap method, using a random re-sampling repeated 10,000 times. In Fig.~\ref{fig:peakscorr}, the black dashed line represents the 100\% correlation between the flaring and non-flaring cases.

From Fig.~\ref{fig:peakscorr}, we can see that the dependence of the EM clearly becomes deviated from that of the SH and T fluxes. Also, the mean periods
of EM of the non-flaring ARs are longer than the flaring ones. These findings strongly suggest that the evolution of the EM flux component has a more prominent role in the flare-CME triggering processes when compared to that of the other two helicity flux components.

Previously, we determined the harmonics for each case of the period peaks of the oscillatory behaviour of the various helicity flux components.
From this, we can see that only the peaks appearing in the EM of flaring ARs are the ones that follow the properties of the harmonics well for an
oscillatory waveguide system. The $1/n$ dependence (n=1,2,3,...) of the fitted
Gaussians of EM of flaring ARs is evidence of such an oscillatory system (Fig.~\ref{fig:fig2}d). Such a clear harmonic property is not detected in the different flux components of non-flaring ARs. 

In summary, if the following series of events/features occur, during the evolution of an AR, these may alert us to an impending flare/CME:
\begin{itemize}
    \item once a $\delta$-spot is forming,
    \item where shorter periods appear in the EM,
    \item where these periods show the properties of harmonics of a resonant waveguide system.
\end{itemize}

\begin{deluxetable*}{cccccc}
\tablecaption{Magnetic helicity flux periods obtained with GMM (5 components) \label{tab:5gmm}}
\tablewidth{0pt}
\tablehead{
\twocolhead{} &
\twocolhead{Flaring active regions} &
\twocolhead{Non-flaring active regions} \\
\colhead{} &
\colhead{Harmonics} &
\colhead{with avg (h)} & 
\colhead{w/o avg (h)} &
\colhead{with avg (h)} &
\colhead{w/o avg (h)}
}
\startdata
                    & f & 19.0 [19.0] & 25.5 [25.5] & 21.0 [21.0] & 21.9 [21.9] \\
Emergence magnetic  & p$_{1}$ & 10.6 [9.5] & 19.5 [12.8] & 17.5 [10.5] & 17.8 [11.0] \\
                    & p$_{2}$ & 5.9 [6.3] & 9.2 [8.5] & 11.0 [7.0] & 10.8 [7.3] \\
helicity flux       & p$_{3}$ & 3.9 [4.8] & 4.9 [6.4] & 4.5 [5.3] & 4.5 [5.5] \\
                    & p$_{4}$ & 2.6 [3.8] & 3.0 [5.1] & 2.7 [4.2] & 2.7 [4.4] \\\hline
                    & f & 19.5 [19.5] & 20.7 [20.7] & 20.5 [20.5] & 18.7 [18.7] \\
Shearing magnetic   & p$_{1}$ & 14.1 [9.8] & 12.2 [10.4] & 13.5 [10.3] & 13.1 [9.4] \\
                    & p$_{2}$ & 7.0 [6.5] & 5.8 [6.9] & 9.2 [6.8] & 9.1 [6.2] \\
helicity flux       & p$_{3}$ & 4.7 [4.9] & 3.3 [5.2] & 5.2 [5.1] & 5.0 [4.7] \\
                    & p$_{4}$ & 3.0 [3.9] & 0.5 [4.1] & 3.1 [4.1] & 3.1 [3.7] \\ \hline
                    & f & 20.3 [20.3] & 20.8 [20.8] & 19.7 [19.7] & 19.4 [19.4] \\
Total magnetic      & p$_{1}$ & 13.1 [10.2] & 13.1 [10.4] & 11.7 [9.9] & 12.8 [9.7] \\
                    & p$_{2}$ & 7.2 [6.8] & 7.7 [6.9] & 6.1 [6.6] & 8.7 [6.5] \\
helicity flux       & p$_{3}$ & 4.5 [5.1] & 4.5 [5.2] & 4.2 [4.9] & 5.2 [4.9] \\
                    & p$_{4}$ & 2.8 [4.1] & 2.8 [4.2] & 2.8 [3.9] & 3.1 [3.9] \\ 
\enddata
\tablecomments{The mean values of the obtained (GMM) Gaussian distributions of the studied distributions are listed. Square brackets are the
fundamental periods with the associated harmonics assuming the system is a uniform resonant waveguide.}
\end{deluxetable*}

\subsection{Distribution of peaks of flaring ARs} \label{sec:flaringARs}

\begin{figure*}
\plotone{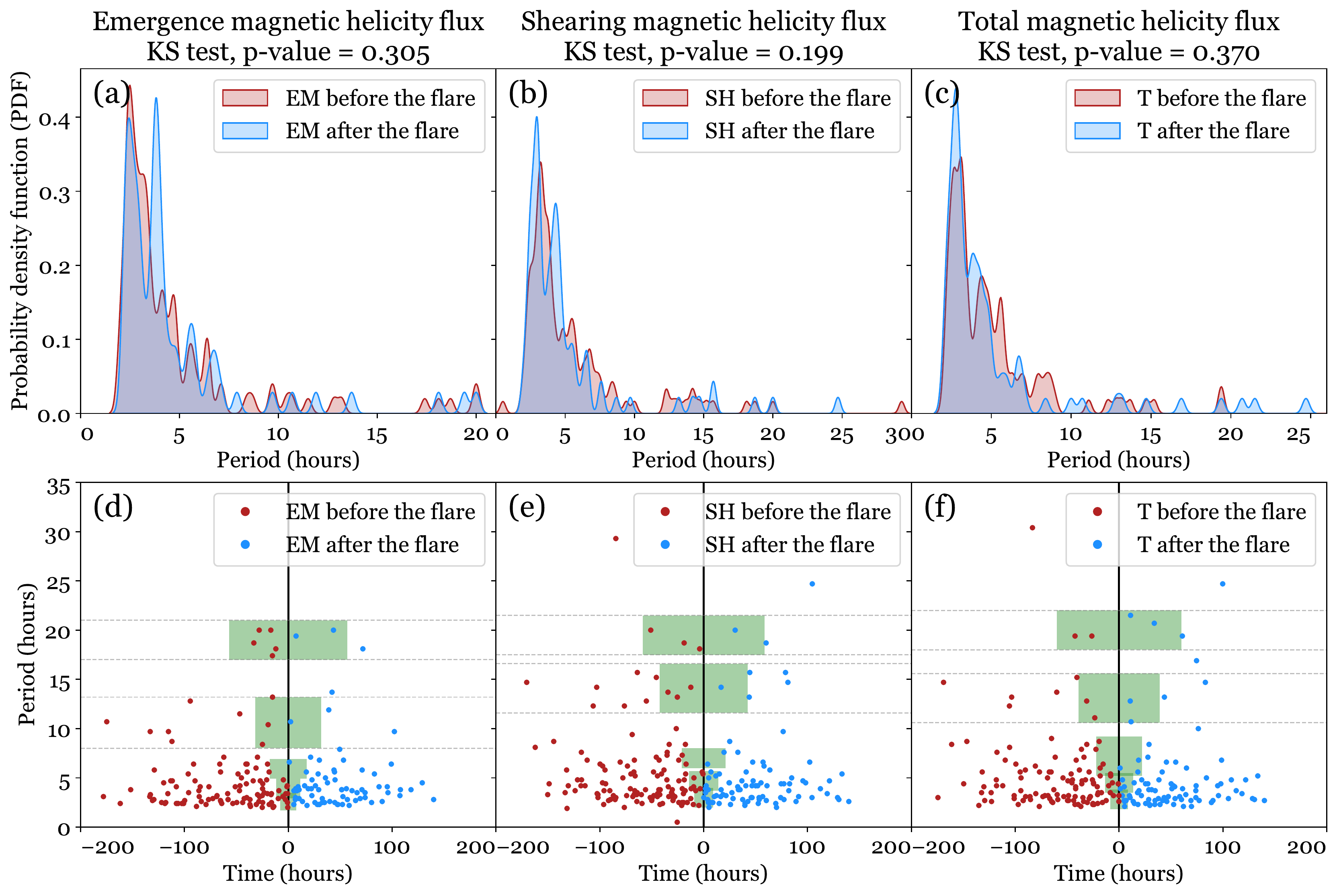}
\caption{ Panels a-c: Distribution of EM/SH/T peaks before (red) and after (blue) the flares. In the EM (a) case the most striking feature is that the
short periods split into $\sim$2.5 and $\sim$4.5 hours after the flare occurred. Similar to the peaks of SH (b). Panels d-f: EM/SH/T periods of all
flaring ARs. The black vertical line marks the reference time of the flares. The center of the green rectangles is given by the center of the
Gaussian distributions fitted by the GMM method. The height of the rectangles corresponds to the FWHM value of the Gaussians fitted by the GMM
method. The width of the green rectangles corresponds to 3 periods. The grey dashed lines are the boundaries of green rectangles, in case of the longer periods of EM/SH/T. \label{fig:fig4}}
\end{figure*}

We now investigate the periodicity distributions of peaks before and after the flares. In Fig.~\ref{fig:fig4}a, we plot the distribution of the localised WPS peaks of EM before (red) and after (blue) the largest intensity flares.
Based on the KS test with \textit{p} = 0.305, the two distributions are different: after the flare, the main peak splits into two distinct peaks ($\sim$2.5 and $\sim$4.5 hours).

In the SH case, (Fig.~\ref{fig:fig4}b), the KS test with \textit{p} = 0.199 indicates that the period distributions are different before and after the flare. The most striking feature here is, again, that the short periods split into two groups after the flare.

The T component behaves similarly to EM, in Fig.~\ref{fig:fig4}c. The magnitudes of the periods shift towards lower values after the flare onset, and, the higher periods mostly disappear. Likewise to EM, the KS test result of \textit{p}-value = 0.370 shows that the different period distributions before and after the flare are not as pronounced as in the SH case.

Next, Figs.~\ref{fig:fig4}d-f show the periodicities as a function of time relative to the onset time of the largest flare. In general, we find that shorter periods (2--8 hours), are continuously present. However, the peak of longer periods ($\ge$10 hrs) are more often observed
before or just a few hours after the flares. 
Interestingly, the peaks of the longer periods of EM/SH/T do not split up into further bands after the flares, as can be seen between the dashed lines in Fig.~\ref{fig:fig4}d-f.


In summary, the shorter periods ($<$10 hrs) seem to separate into two groups after the flare (see Figs.~\ref{fig:fig4}a-c), while 
the longer periods do not (see Figs.~\ref{fig:fig4}d-f).

\subsection{Distribution of peaks of non-flaring ARs} \label{sec:nonflaringARs}

Next, the peak distributions of non-flaring ARs are examined, separately for the EM, SH and T helicity fluxes. Since there is no set moment of flare onset time in this case, we select an arbitrary reference time in every non-flaring AR. Since the average investigated time interval of an AR was about 7-9 days in duration, we define a set of reference times ranging from 0 to 200 hrs, in 5-hour increments (see the three corresponding animations for EM, SH, and T in the online material). This range of about a total of 40 different reference times helps to avoid bias in the analysis ({\it Note: Please find the animations in the Appendix.}).

For each arbitrarily chosen time, the distributions of EM/SH/T are analysed. The studied time intervals (between 50 and 140 hours) give an 
appropriate distribution before and after the arbitrary time, because outside of these time intervals, the peaks sometimes run very high due to the
normalization of the distribution. For this reason, we filter the animations and cut off the beginning and end of the arbitrarily chosen time intervals.

In summary, and most importantly, there is no significant change in the peak distribution before and after a suitably chosen (say 40--50 to 130--145 hour)
arbitrary time when compared to the corresponding counter-parts of analyses of flaring ARs. Indeed, this is the expected behaviour since there is no naturally distinguished physical reference time in non-flaring ARs when the conditions of helicity oscillations may undergo rapid changes,
as occurs in flaring ARs. Overall, from the results found so far, we may safely conclude that there is an intrinsic relationship between the
periodic oscillations of the helicity fluxes and the flaring activities of an AR.

\subsection{Comparing the number of periods in ARs} \label{sec:flaringvsnonflaring}

In Figs.~\ref{fig:fig4}d--\ref{fig:fig4}f, the periods of WPS peaks of all flaring ARs are plotted, separately for the EM, SH, and T fluxes. The black vertical
lines represent the reference time of each flare event. The center of the green
rectangles is determined by the center of the Gaussian distributions fitted by the GMM method (as seen in Table~\ref{tab:5gmm}). The height of the
rectangles corresponds to the FWHM value of the Gaussians, and the width of the rectangles corresponds to 3 whole periods. This indicative width is chosen because by multiplying the means of Gaussians by $e$-fold, one would have about 3 oscillatory periods.

\begin{figure*}
\plotone{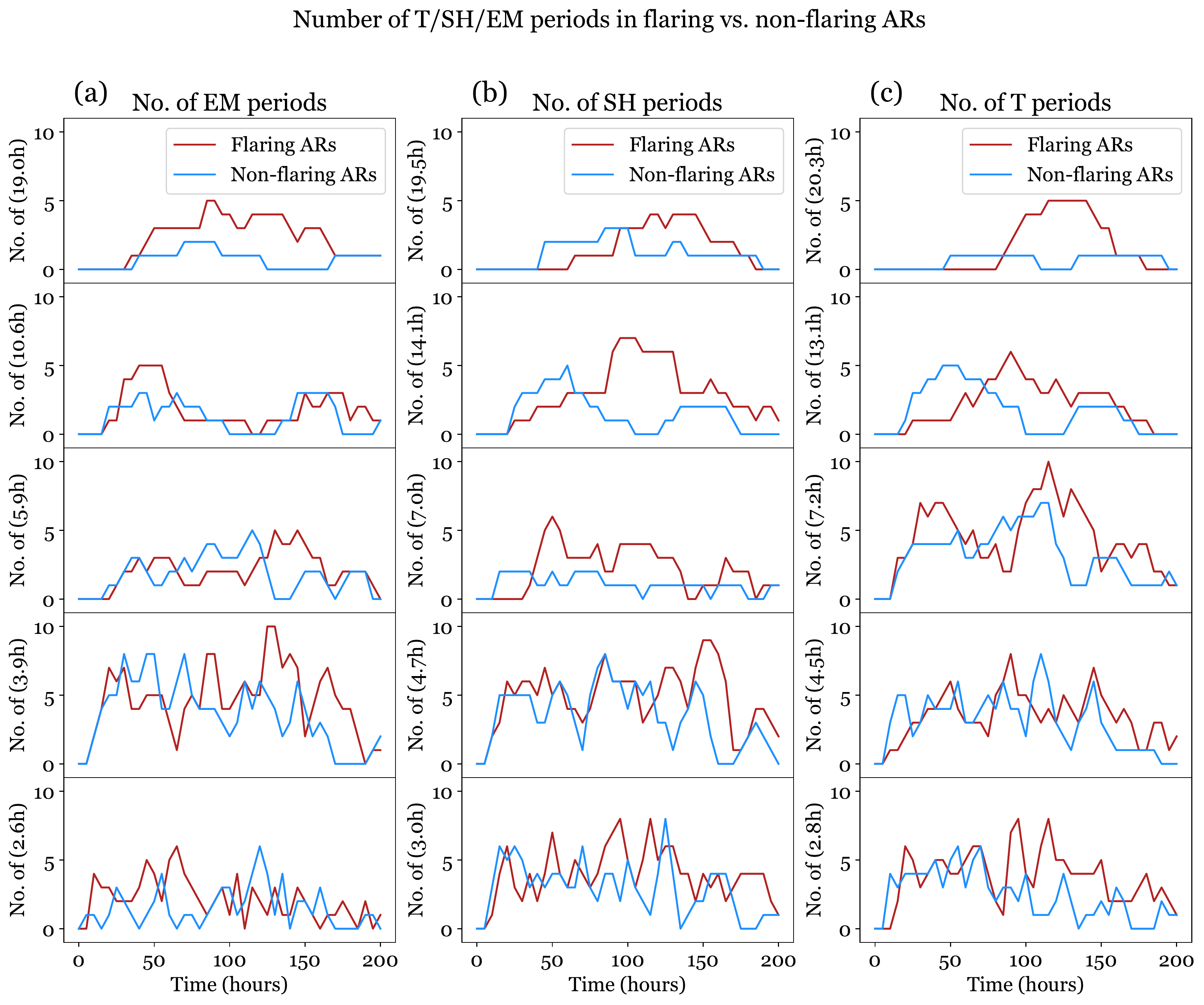}
\caption{Number of (a) EM, (b) SH and (c) T oscillatory peak periods in flaring and non-flaring ARs. The blue/red lines correspond to non-flaring/flaring ARs. Long periods ($\sim$20 hour) appear in the oscillatory peaks of the WPS of the T and EM fluxes of flaring ARs, which seem to be linked to flare/CME eruptions. \label{fig:totshemh}}
\end{figure*}

We count the number of peaks appearing within each defined rectangular region for the flaring ARs, and for an interval three periods earlier of the
arbitrarily chosen reference times for the non-flaring ARs. In Fig.~\ref{fig:totshemh}, we plot the number of points summed from each green rectangle as
a function of the arbitrarily chosen reference times for the magnetic helicity flux components of flaring and non-flaring ARs. The
\textit{x}-axis of Fig.~\ref{fig:totshemh} corresponds to the non-flaring ARs reference times ranging from 0 to 200 hrs, in 5-hour increments. A total of 40 different
reference times as in Section \ref{sec:nonflaringARs}. For the EM case (Fig.~\ref{fig:totshemh}a), long ($\sim$19-hour) periods are present throughout,
and are significantly higher for the flaring ARs. For shorter periods, it is difficult to distinguish between flaring and non-flaring ARs.
In the SH case (Fig.~\ref{fig:totshemh}b), the maximum count is clearly identifiable at $\sim$14.1-hour. For the flaring ARs, the long periods ($\sim$20-hour)
appear for the total helicity flux T (Fig.~\ref{fig:totshemh}c), from which we conclude that if these periods appear, we may expect flare/CME eruptions.
No such clear difference can be established from the shorter periods. At the moment of $\sim$13.1 hour, the number of flaring periods is slightly higher
than in the non-flaring case, but this alone is not sufficient to use it for flare warning.

The presence of long periods in EM and T suggests that it plays a crucial role
in the formation of flares.

\section{Summary} \label{sec:summary}

In this work, we tested and further developed the results found in \citet{Korsos_2020}, by carrying out wavelet analysis, about the dynamic evolution of emergence (EM), shearing (SH), and total (T) magnetic helicity flux terms. \citet{Korsos_2020} reported a unique relationship between the oscillatory behavior of the three magnetic helicity flux components and the associated flare activities.

To test their conjecture, here, we have analyzed the EM, SH, and T magnetic helicity flux evolution of 14 flaring and 14 non-flaring ARs. 
Following the methodology of \citet{Korsos_2020}, first we mitigated the artificial 12- and 24-hr periods of the SDO/HMI magnetogram measurements by set lower $\left| \pm 200\right|$ G and upper $\left| \pm 2000\right|$ G magnetic field boundaries for an AR, based on \cite{Smirnova_2013} and \cite{Tziotziou_2015}. To further reduce the 12- and 24-hr SDO artifacts, we smoothed the time series of EM/SH/T with 24-hr smoothing window and subtracted the obtained averaging from the original data.

As a next step, we have constructed the wavelet power spectrum (WPS) of EM/SH/T time series. Regions of the WPS at above the 1$\sigma$ significance level were identified and the peaks contained in these regions were recorded. 
Before finalising the $\left| \pm 200\right|$ G lower and $\left| \pm 2000\right|$ G upper boundaries, we also extensively looked into how these boundaries change the results when limiting the magnetic field strength. We found that, in general, if one caps the magnetic field strength, the lower (3–15 hrs) periods become less frequent, like it was suggested by \cite{Smirnova_2013}.

After determining the lower and upper magnetic field boundary values, the following results were found by means of statistical analysis (see for more details Sec.~\ref{sec:analysis}) of the identified local peaks within 1$\sigma$ significance level of the corresponding WPS:
\begin{itemize}
    \item For flaring and non-flaring ARs, the EM/SH/T periodicities occur in bands. These bands are between (i) 2--9 hours, (ii) 11--14 hours, and (iii) 19--21 hours (see Figs.~\ref{fig:fig2}d-f).
    
    \item The distribution of EM/SH/T peak periodicities were fitted using a Gaussian Mixture Model. Fig.~\ref{fig:fig2}d shows the $1/n$-dependence ($n=1,2, 3, ...$, a positive integer) of the GMM-fitted EM peaks for flaring ARs. However, such clear harmonic oscillatory properties were not present in the SH/T flux components of flaring and in the EM/SH/T fluxes of non-flaring ARs.

    \item 
    There is a noticeable difference in the distribution of central periods found in the EM profiles between the flaring and non-flaring ARs, while no significant difference is found in the cases of the SH and T profiles.
    The central periodicities of the non-flaring EM are significantly longer than the flaring.
    
   
   \item For the three helicity components, the distribution of lower periods ($<$10 hrs) are concentrated around $\sim$2.5 hrs before the flare events, see Fig.~\ref{fig:fig4}a-c. Interestingly, these lower periods are separated into two groups (i.e. $\sim$2.5 and $\sim$4 hours), after the flares (see Fig.~\ref{fig:fig4}a-c). This could be explained by the re-arrangement and disappearance of the magnetic field as a waveguide resonator “allowing” oscillations around only $\sim$2.5 hour prior flaring. However, the distribution of the longer periods ($>$10 hrs) does not change after the flares, see the highlighted areas with dashed lines in Fig.~\ref{fig:fig4}d-f.

    
    \item In the flaring AR cases, the stronger presence of long periods in the EM (i.e. $\sim$19 hour) and in T ($\sim$20 hour) oscillatory data would suggest that the EM component does play a more crucial role in the formation of flares (see Fig.~\ref{fig:totshemh}). This condition is only indicative as these periods may also appear in few cases of non-flaring ARs (see the top panels in columns a and c of Fig.~\ref{fig:totshemh}).
    
    \item In the non-flaring ARs studied, there is no significant change in the peak distributions before and after a suitably chosen (say 40--50 to 130--145 hour) arbitrary reference time when compared to the counterpart distributions before and after in flaring ARs (see Figs.~\ref{fig:a1}, \ref{fig:a2} and \ref{fig:a3}).
    
    \item To test the robustness of our findings, we look for a bias in the selection of AR samples in two ways: i) we exclude a random AR from the 14 flaring AR samples, as well as one from the 14 non-flaring AR samples; and ii) we generate synthetic samples using bootstrap method, then we reconstruct Figs.~\ref{fig:fig2}, \ref{fig:peakscorr}, \ref{fig:fig4} and \ref{fig:totshemh} again. In both cases we came to the same conclusion that neither the synthetic samples nor the exclusion of 1-1 ARs change the results significantly.
\end{itemize}

Based on the above, we conclude that there is an intrinsic relationship between the periodic oscillations of the helicity fluxes and flare activity. Our results show that the evolution of the EM helicity flux component has a more prominent role in the flare-CME triggering process, especially when 
\begin{itemize}
    \item the AR has a $\delta$-spot,
    \item the shorter oscillatory periods appear in the EM flux data, and
    \item these periods show the presence of a harmonic oscillatory resonator.
\end{itemize}
To apply these results as a precursor, or, use them to distinguishing between flaring and non-flaring ARs, we need at least $\sim$2.5 days data to study the evolution of the three helicity flux components and reveal the possible characteristic long period(s) of a flaring AR by wavelet analyses.

\software{MUDPACK \citep{ADAMS_1993},
          Wavelet Analysis \citep{Torrence_1998},
          Persistent Homology \citep{Huber_2021},
          seaborn \citep{Waskom2021},
          sklearn \citep{scikit-learn},
          scipy \citep{SciPy}}

\section*{Acknowledgements}
The authors thank the referee and the handling editor for their constructive comments that have improved the paper. The authors also acknowledge the support received from OTKA (grant number K128384), Hungary. MBK and HM are grateful to the Science and Technology Facilities Council (STFC), (UK, Aberystwyth University, grant number ST/S000518/1), for the support received while carrying out this research. RE is grateful to STFC (UK, grant number ST/M000826/1) and EU H2020 (SOLARNET, grant number 158538).
RE also acknowledges support from the Chinese Academy of Sciences President’s International Fellowship Initiative (PIFI, grant number 2019VMA0052) and
The Royal Society (grant nr IE161153).

\appendix
\label{sec:gifs}

Here, we list the associated animations of the distribution of peaks of non-flaring ARs, available in the online version. The animations show the
distribution of periods before the arbitrarily chosen reference moment of time (in blue) and the distribution of periods after the arbitrarily
chosen reference time (in red).

\begin{figure}
\begin{interactive}{animation}{f8.mp4}
\plotone{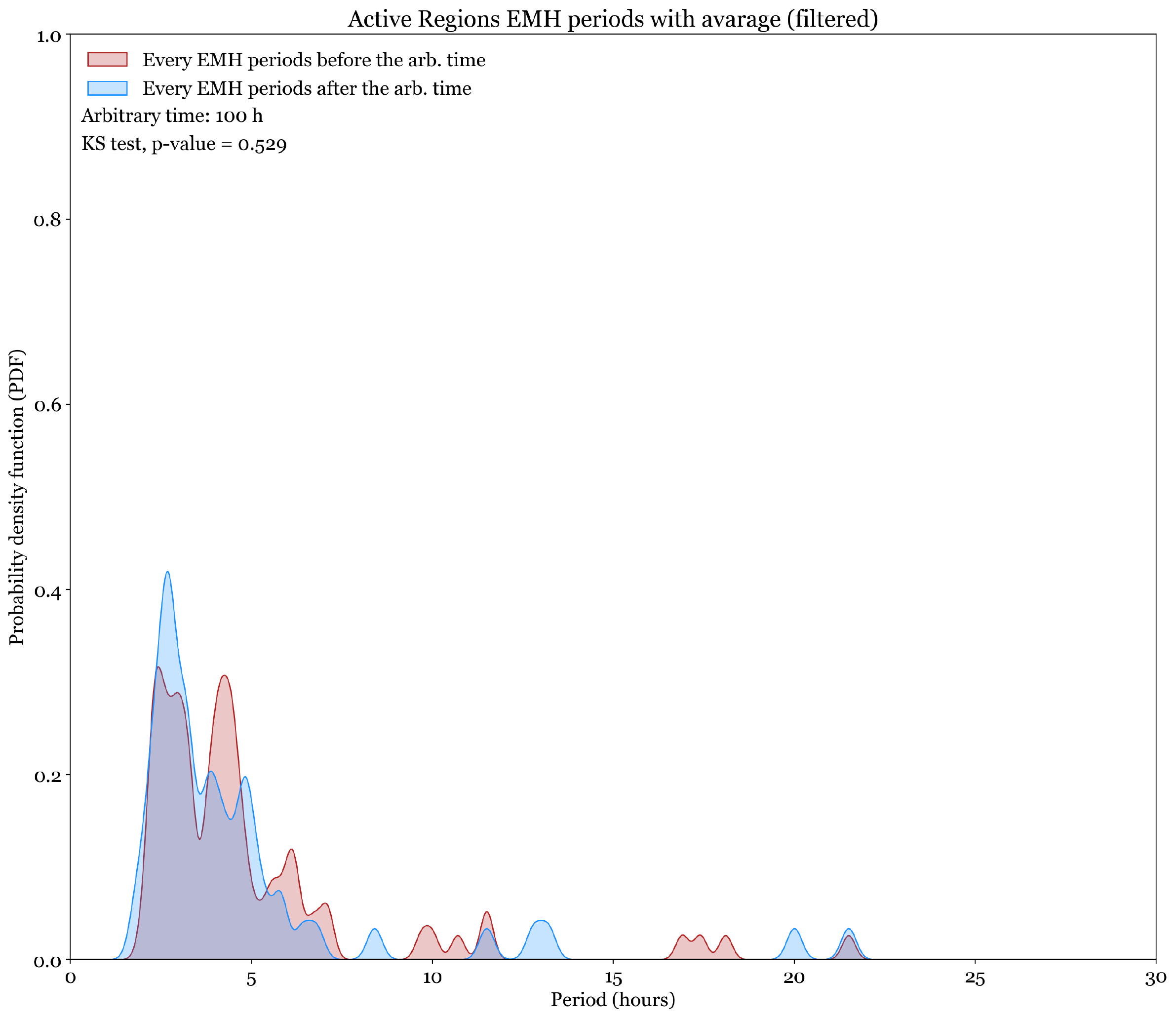}
\end{interactive}
\caption{This figure is available as an animation which shows consecutive reference times of non-flaring EM before/after the chosen reference time. Only small fluctuations are visible. The animation runs between arbitrary times 40 - 140 hours. The real-time duration of the animation is 2 seconds.\label{fig:a1}}
\end{figure}

\begin{figure}
\begin{interactive}{animation}{f9.mp4}
\plotone{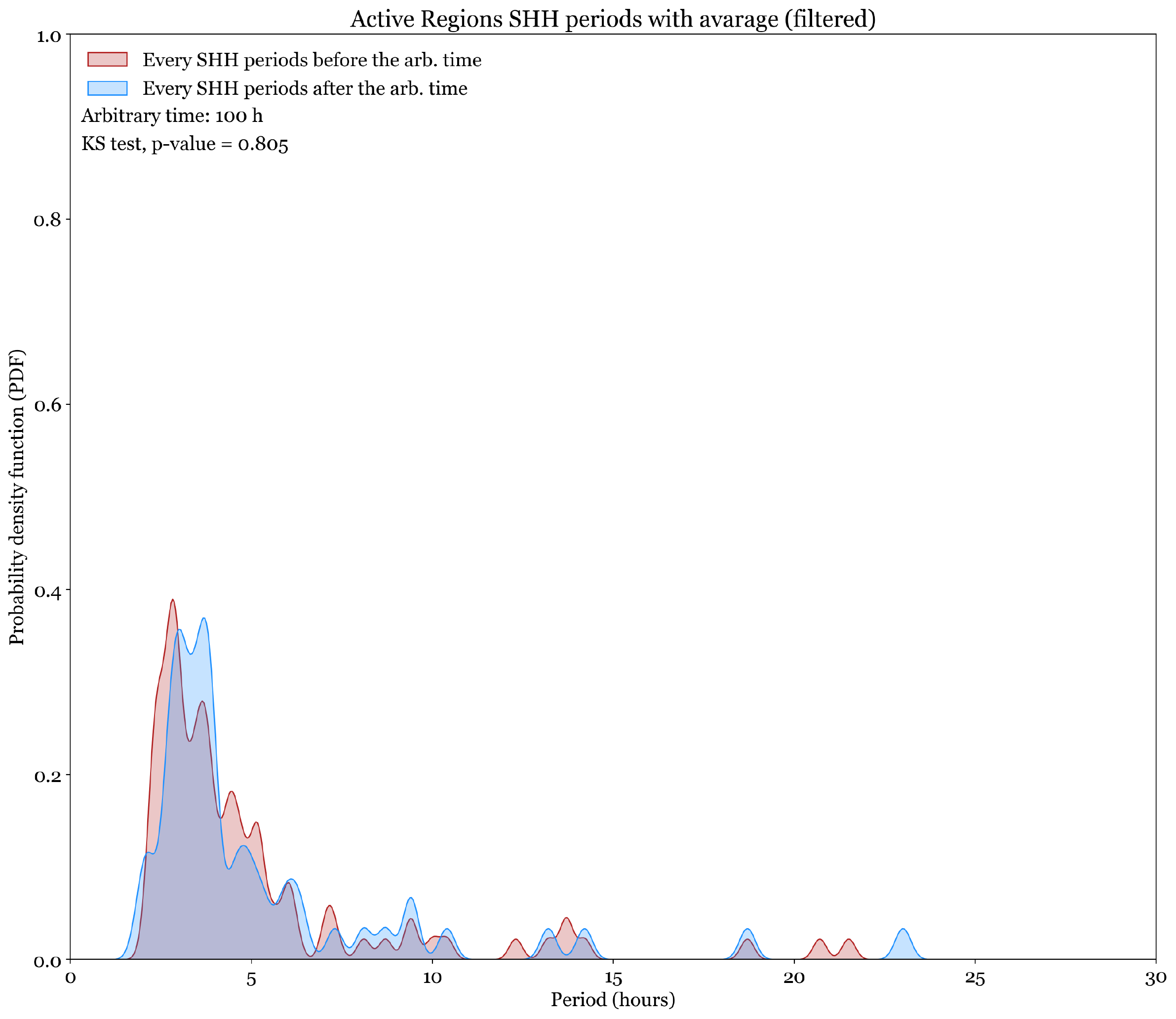}
\end{interactive}
\caption{This figure is available as an animation which shows consecutive reference times of non-flaring SH before/after the chosen reference time. Only small fluctuations are visible. The animation runs between arbitrary times 45 - 130 hours. The real-time duration of the animation is 2 seconds.\label{fig:a2}}
\end{figure}

\begin{figure}
\begin{interactive}{animation}{f10.mp4}
\plotone{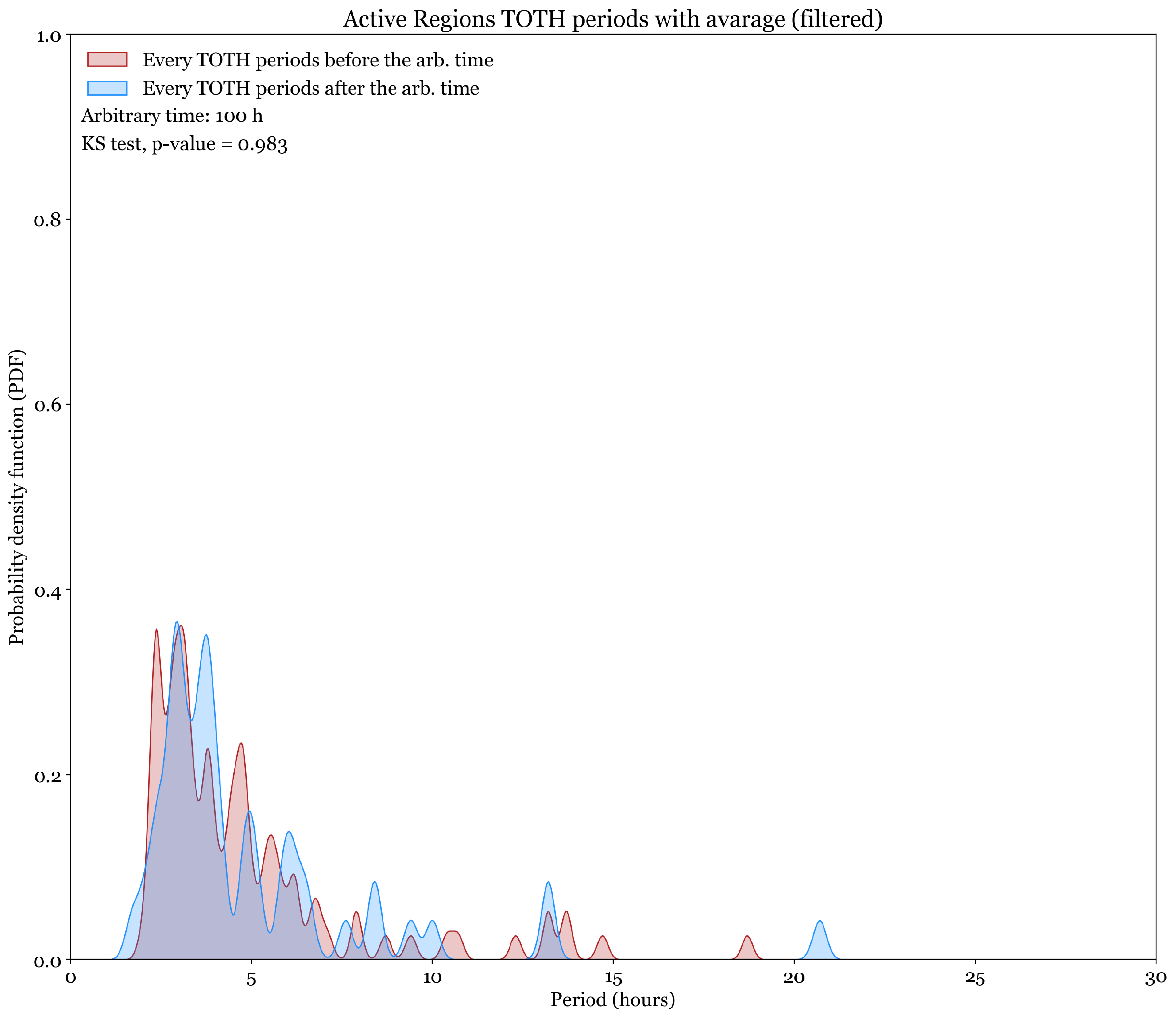}
\end{interactive}
\caption{This figure is available as an animation which shows consecutive reference times of non-flaring T before/after the chosen reference time. Only small fluctuations are visible. The animation runs between arbitrary times 50 - 125 hours. The real-time duration of the animation is 2 seconds.\label{fig:a3}}
\end{figure}

\bibliography{soos_paper_1}{}
\bibliographystyle{aasjournal}

\end{document}